\documentclass[conference]{IEEEtran}
\usepackage[numbers]{natbib}

\usepackage{balance}
\usepackage[italic]{mathastext}
\usepackage{amsmath,amssymb,amsfonts}
\usepackage{algorithmic}
\usepackage{graphicx}
\usepackage{textcomp}
\usepackage{xcolor}
\usepackage{hyperref}
\usepackage{multirow}
\usepackage{enumitem}

\usepackage{booktabs}

\def\BibTeX{{\rm B\kern-.05em{\sc i\kern-.025em b}\kern-.08em
    T\kern-.1667em\lower.7ex\hbox{E}\kern-.125emX}}

\newcommand{\rr}[1]{\textcolor{black}{#1}}

\newcommand{\pid}[1]{\textit{{#1}}}
\newcommand{\inlineq}[1]{\textit{``#1''}}
\newcommand{\inlineqq}[2]{\textit{``#1''{(#2)}}}

\begin{document}

\title{Designing Telepresence Robots to \\ Support Place Attachment}
% \title{Multi-party Remote Experience through Telepresence Robots toward Place Attachment}

% {\footnotesize \textsuperscript{*}Note: Sub-titles are not captured in Xplore and
% should not be used}

% \thanks{Identify applicable funding agency here. If none, delete this.}

% \author{\IEEEauthorblockN{\textit{Submission ID:} 1127}}
\makeatletter
\newcommand{\linebreakand}{%
  \end{@IEEEauthorhalign}
  \hfill\mbox{}\par
  \mbox{}\hfill\begin{@IEEEauthorhalign}
}
\makeatother
\author{

\IEEEauthorblockN{\href{https://orcid.org/0000-0003-4462-0140}{Yaxin Hu}}
\IEEEauthorblockA{\textit{Department of Computer Sciences} \\
\textit{University of Wisconsin--Madison}\\
Madison, Wisconsin, USA \\
\href{mailto:yaxin.hu@wisc.edu}{yaxin.hu@wisc.edu}
% \\
% \href{https://orcid.org/0000-0003-4462-0140}{0000-0003-4462-0140}
}
\and
\IEEEauthorblockN{\href{https://orcid.org/0009-0001-9583-8900}{Anjun Zhu}}
\IEEEauthorblockA{\textit{Department of Computer Sciences} \\
\textit{University of Wisconsin--Madison}\\
Madison, Wisconsin, USA \\
\href{mailto:azhu39@wisc.edu}{azhu39@wisc.edu}
% \\
% \href{https://orcid.org/0009-0001-9583-8900}{0009-0001-9583-8900}
} 
\linebreakand
\IEEEauthorblockN{\href{https://orcid.org/0000-0003-0714-312X}{Catalina L. Toma}}
\IEEEauthorblockA{\textit{Department of Communication Arts} \\
\textit{University of Wisconsin--Madison}\\
Madison, Wisconsin, USA \\
\href{mailto:ctoma@wisc.edu}{ctoma@wisc.edu}
% \\
% \href{https://orcid.org/0000-0003-0714-312X}{0000-0003-0714-312X}
}
\and
\IEEEauthorblockN{\href{https://orcid.org/0000-0002-9456-1495}{Bilge Mutlu}}
\IEEEauthorblockA{\textit{Department of Computer Sciences} \\
\textit{University of Wisconsin--Madison}\\
Madison, Wisconsin, USA \\
\href{mailto:bilge@cs.wisc.edu}{bilge@cs.wisc.edu}
% \\
% \href{https://orcid.org/0000-0002-9456-1495}{0000-0002-9456-1495}
}
}

\maketitle
% - People want to remain connected to places that are meaningful to them.
% - This is challenging due to travel, mobility issues, etc.
% - Telepresence robots can help.
% - How should these robots be designed to best provide this help? (how do you deal with multiple parties, can the robot be autonomous, design decisions, etc. 
% - In this paper, we explore ....
\begin{abstract}
    People feel attached to places that are meaningful to them, which psychological research calls ``place attachment.'' Place attachment is associated with self-identity, self-continuity, and psychological well-being. Even small cues, including videos, images, sounds, and scents, can facilitate feelings of connection and belonging to a place. Telepresence robots that allow people to see, hear, and interact with a remote place have the potential to establish and maintain a connection with places and support place attachment. In this paper, we explore the design space of robotic telepresence to promote place attachment, including how users might be guided in a remote place and whether they experience the environment individually or with others. We prototyped a telepresence robot that allows one or more remote users to visit a place and be guided by a local human guide or a conversational agent. Participants were 38 university alumni who visited their \textit{alma mater} via the telepresence robot. Our findings uncovered four distinct user personas in the remote experience and highlighted the need for social participation to enhance place attachment. We generated design implications for future telepresence robot design to support people’s connections with places of personal significance.

% People form attachment to places that are meaningful for them and such place attachment is associated with their perceived self-identity, self-continuity, and psychological well-being. For people who can not access these places in person, videos, images, sounds and scents can facility the feeling of connection and belonging to these places. Telepresence robots that allow people to see, hear, and interact with the remote place has the potential to facilitate the feeling of place attachments. 

% In this paper, we studied the use of multi-party telepresence robots to facilitate the remote users' interaction with places with significant personal meanings. We prototyped a telepresence robot that allows one or multiple remote users to visit a place together, with a local guide user or an conversational agent. We focused on a scenario where alumni remotely visit their alma mater through the robot. Through field studies with 38 alumni participants, we studied four configurations of the mutli-party interactions with the telepresence robot for the place engagement. Our findings highlighted X Y Z. We drew design implications for the multi-party system for future user engagement with places and create meaningful experiences.
\end{abstract}

% comment sep 17
% telepresence robot can connect with place - how are we going to do that --> where is the guide from? --> the study design clearly follows the design space  --> then we say we have a local guide and agent guide 


\section{Introduction}
Place attachment is a phenomenon in which people form emotional attachment to physical environments~\cite{inalhan2004place}. Attachment to a place is associated with a greater sense of coherence, higher life satisfaction, stronger social bonds and neighborhood ties, greater interest in family roots, greater trust in others and less egocentrism~\cite{lewicka2011varieties}. Disruptions in connection to a place, for example due to relocation or disasters, introduce a significant amount of stress, threaten self-identity, and trigger a process of coping with lost attachments and seeking new ones~\cite{brown1992disruptions}. After moving away, to avoid a loss of self-identity, people seek to maintain connection with places by physically visiting them~\cite{li2016effects} or remotely experiencing them via computer-mediated communication (CMC) technologies~\cite{hiller2004new,gustafson2014place}.

\begin{figure}[tb]
    \centering
    \includegraphics[width=\linewidth]{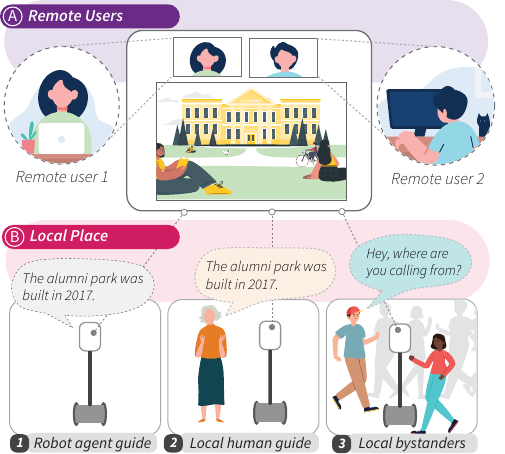}
    \caption{In this paper, we explore the design space of the use of telepresence robots to support place attachment, specifically how robots might accommodate multiple users (A) to remotely experience places that are meaningful to them for a range of scenarios (B).}
    \label{fig:teaser}
    \vspace{-12pt}
\end{figure}

\textit{Telepresence robots} are one such technology that holds significant promise in maintaining ties with places that are meaningful to people. Early visions and prototypes of telepresence robots explored the potential for people to physically interact with remote environments~\cite{paulos1997ubiquitous,paulos2001personal}. Applications of telepresence robots developed henceforth focused on facilitating presence and social interaction in specific environments, including allowing students to attend school and other educational activities~\cite{lei2022acceptance, elmimouni2024navigating}; helping people access health services, such as postoperative care~\cite{kristoffersson2013review}; allowing participation in teamwork and hallway conversations in work environments~\cite{tsui2011telepresence}; and social connection-making for older adults~\cite{beer2011mobile,hung2022facilitators} and individuals with dementia~\cite{moyle2017potential}. In these contexts, studies have found robotic telepresence to provide users with a greater sense of presence~\cite{rae2014bodies} and alleviate the negative effects of distance on communication~\cite{kornfield2021so} compared to video-based telepresence. Despite the wide-ranging applications of telepresence robots across many domains and the potential benefits of robotic telepresence, little is known about how this technology can support place attachment.

This paper investigates the use of telepresence robots to support place attachment after disruptions, and explores the design space of such systems, focusing on how remote users might be guided in the place and whether users experience the place individually or with others. Specifically, we address the following research questions. $RQ_1$: \textit{How do individuals or groups use telepresence robots to connect with a place?} $RQ_2$: \textit{How can telepresence robots facilitate human-place connection?} To address these questions, we prototyped a telepresence robot system, using the Double 3 telepresence robot as a platform. We included the ability to support multiple users and provided guidance by a local guide or a conversational agent controlled by a mixture of large language model (LLM) and Wizard of Oz (WoZ) approach. 

To study use patterns and effects of the guide type and the number of remote users, we deployed the robot in key locations on a university campus and recruited university alumni ($n=38$) to visit locations of their choice via the robot. Our findings highlighted place attachment experiences on the personal, group and community levels, and identified four personas of visitors. We discuss implications for the potential of telepresence robots to support place attachment, as well as for the design of future robot systems.
% $RQ_1$: \textit{How do people who use telepresence robots to connect with a place?} 
% $RQ_1$: \textit{How do individuals and groups differ in their experiences of places via a telepresence robot?} 
Our work makes the following contributions:

\begin{itemize}
  \item \textit{Design \& Artifact Contributions:} We designed and prototyped a telepresence robot that can provide human/agent-based guidance to one or multiple remote users.
  \item \textit{Empirical Contributions:} We conducted a field study to understand use patterns and how different design factors affect these patterns and user experience.
  \item \textit{Practical Contributions:} We discuss opportunities for future design and provide specific guidelines to support place attachment.
\end{itemize}

% Prior research has explored the use of telepresence robot

% What is the motivation of the paper?

% What has been done? 
% What are the related work and related field? 

% What is the gap? 

% What is your approach? Why is this approach? 
% How is your approach different from the others?

% What are your research questions?
% What did you do to answer the research questions?
% What was your findings?
% What was the implications for your findings?

% Contributions:

% Research questions: 
% - frequency based - what are the interactions looking like?
% - qualitative - people's feeling
% -- when do I present the freq vs open end 

% ========== Multi-party Interaction Pattern ========== 
% What are the interaction patterns for each configuration of the multi-party interaction?
% What types of activities occur during the remote experience? 
% What multi-party interactions were preferred? What people's feedback for each configuration?
% -- With the guide
% -- With the robot agent
% -- Two visitors 

% What are people's feedback for the remote experience? 
% How are their feedback different across the four conditions?

% ========== Meaningful Experiences with the Place ========== 
% How can telepresence enhance person-place bonding? 
% - What are the activities? 
% - What emotions are triggered? 
% - What are the factors involved? 
% What are the meaning-making process in the interaction? What new meanings were created? How has this been beneficial for the users' psychological wellbeing? 

\section{Related Work}
% Place attachment theory
% Multi-party Telepresence Robots
% Conversational Telepresence Robots
% Technology for meaningful experiences / place attachment 
% \todo{We thank our reviewers for suggesting additional literature to enhance our related work (R2, R3) and replace weaker references (R3).
% }
\subsection{Place Attachment Theory}
Place attachment is a psychological process where people naturally form bonding with places with significant personal meanings~\cite{tuan1979space, lewicka2011place, low1992place}. A person's sense of a place is formed through the their socialization with the physical world~\cite{twigger1996place}. Place attachment is a multifaceted concept. It has been studied in the personal, community, and natural environment contexts~\cite{raymond2010measurement}. On the personal level, place attachment is associated with a sense of safety and security, self-continuity, self-esteem, and self-identity. When comparing the past and the present in a particular place, one can enhance a sense of self-continuity and reinforce personal identity~\cite{scannell2010ppp, scannell2017experienced}. On the community level, place attachment is associated with social bonding with and attachment to the local community, and a sense of belonging to a social group~\cite{kasarda1974community, kyle2007social, cresswell2014place}. On the environment level, place attachment is related to an affinity and connection with nature, and is closely related to leisure activities conducted in various nature settings~\cite{kyle2004examination}. Place attachment is a continuous and dynamic process~\cite{inalhan2004place}. The past place becomes an archive of memories that inspire the present~\cite{nora1989between, casey2000remembering}. Place attachment can provide a sense of self-continuity~\cite{sedikides2015nostalgia} and meaning of life~\cite{routledge2011past}, contributing to psychological well-being.

\subsection{Technology, Place, and Memory}
% Technologies have been studied to strengthen human-place connection towards people's psychological well-being. 
Prior work has investigated how technologies facilitate the remembering and recalling of place-specific memories and evoke affective experience of the place~\cite{ozkul2015record, kim2022slide2remember, mcgookin2019reveal, white2023memory, sabie2020memory}. \citet{ozkul2015record} studied how sharing location information (\textit{e.g.}, geo-tagging) in mobile media contributes to preserving past memories, creating narratives of the place, and reflecting on the sense of connection between the past, present, and future. \citet{kim2022slide2remember} developed a wall photo frame for people who see their past digital photos and hear songs they had listened to when the photo was taken. 

% \citet{mcgookin2019reveal} studied how a proactive location-based tool can encourage people to revisit their past photos and support reminiscing. \citet{white2023memory} utilized personal location histories to support people's everyday reminisence experience. \citet{nair2006alumni} built a location-based online tour guide system that allow alumni to understand changes of their alma mater.  

% Mobile technologies have been widely studied for people's sense of a place. \citet{ozkul2014location} studied how locative media foster people's attachment to places through maintaining social relations, establishing new social ties, and creating new interests in places. 

\subsection{Telepresence Robots for Place Experience}
Telepresence robots have been widely studied as facilitators of social activities~\cite{tsui2011exploring}, community events~\cite{rae2017robotic, neustaedter2016beam, uriu2021generating}, education~\cite{cha2017designing, tanaka2014telepresence, hu2024telepresence}, elderly care~\cite{hung2022facilitators, stuck_understanding_2017}, and instrumental activities of daily living~\cite{yang2018shopping}. Prior work highlighted the utility of telepresence robots for experiencing remote locations, including visiting museums and cultural heritage sites~\cite{ng_cloud_2015, burgard1999experiences, thrun1999minerva, tsui_accessible_2015}, exploring urban parks~\cite{heshmat2018geocaching} and botanical gardens~\cite{hu2024really}, and going to concerts and sporting events~\cite{beer_mobile_2011}. Grounded in this literature, we emphasized the design opportunity of telepresence robots for human-place bonding and designed a multi-party HRI experience to connect people with places of personal significance. We focused on the scenario of alumni revisiting their \textit{alma mater} and studied how the remote experience through the robot contributed to the sense of place attachment and self-continuity.

\begin{figure}[!tb]
    \centering
    \includegraphics[width=\linewidth]{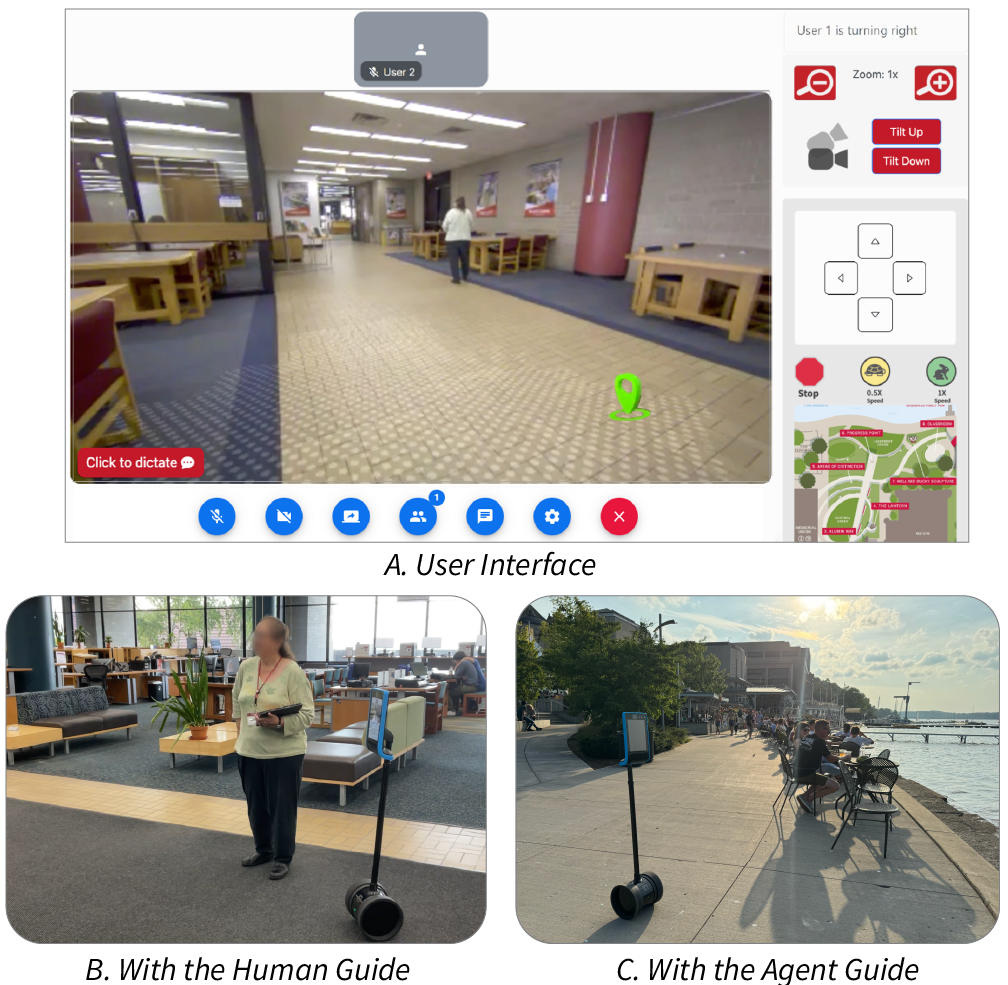}
    \caption{Our prototype system. (A) The interface for remote users to view the local environment, control the robot, and interact with the agent guide and other users. (B) The robot with the human guide in the university library. (C) The robot with the agent guide in the lakefront park.}
    \label{fig:interface}
\vspace{-12pt}    
\end{figure}

% \todo{Improve Figure 2 and Figure 5}

% What is the artifact?
% [Interaction Design] How users interact with it? 
% [System Implementation] How is it implemented?  
% If software architecture, what are the modules and what are they for? 
% What tools/APIs/libraries have you used if using external libraries? 
\section{Design and Prototyping} \label{sec:system design}

\subsection{Design Space} 
We explored the design space for supporting place attachment through telepresence robots based on the social and physical aspects of a place~\cite{scannell2010defining, raymond2010measurement}. We focused on two design factors--local guidance and interaction parties--that provide information about the place and support social participation in the place. For local guidance, we explored the set-up where the user can visit with a local human guide or with an agent guide. The guides can help users control the robot, answer user questions about the place, make suggestions on where to go, and ask users questions about their past experience about the place. We also explored the robot design for multiple remote users to join the experience and control the robot together. Below, we detail our prototype design and implementation. 

% On the personal level, place attachment is associated with people's personal experience, social ties, and memories in the place~\cite{raymond2010measurement}. On the community level, one's attachment with the place is associated with their sense of belonging to the community~\cite{kasarda1974community, kyle2007social, cresswell2014place}. Therefore, w

\subsection{Artifact} We utilized a commercially available telepresence robot platform\footnote{Double 3 telepresence robot: \url{https://www.doublerobotics.com/}} equipped with a screen and cameras for teleconferencing calls, a mobile base with depth sensors, ultrasonic range finders, and an inertial measurement unit (IMU) for obstacle avoidance and semi-autonomous navigation. The robot moves at a speed similar to a moderate walking pace. Users interact with the robot through the control panel on the Web interface or using dialogue. The robot was designed to handle both individual and multi-user interactions. 

\subsection{Interface Design}
The user interface was designed to allow a remote user to join a teleconferencing video call to control the robot and access the robot's environment. The interface was hosted on a website browser and consisted of three components as illustrated in Figure \ref{fig:interface}: (1) a teleconferencing video call module; (2) a robot control panel; and (3) a dialogue interface with the robot. The teleconferencing module allowed multiple remote users to simultaneously join a call with the robot. Users can see a live video feed and hear sounds from the robot's environment. Users can choose to show or hide their face, mute or unmute in the video call with the robot. Users can also adjust the robot camera by zooming in/out and tilting up/down, control the robot's navigation either by pressing the directional buttons on the web interface or using keyboard keys, or placing a robot waypoint in the video feed to have the robot autonomously navigate to a destination. The dialogue interface allowed conversational interactions with the robot. The user can press the dictation button to speak to the robot. Text of speech from the user and the robot was displayed as overlay speech bubbles above the teleconferencing interface. People nearby can also hear the user's dialogue with the robot unless the user was muted. 

 % Users can choose to turn on/off their camera to show or hide their own video feed during the call. 
% its semi-autonomous mode. For manual control, users can drive the robot to move forward/backward, or turn left/right by pressing the D-pad on the interface or using the arrow keys in their keyboard. When using the semi-autonomous control, the user can set a destination point around the robot on the video stream from the robot. 
% For the dialogue button, the user can press it and then dictate their speech to involve in the dialogue with the robot. Both the users' dictated speech and the robot's response are displayed as message bubbles above the video stream in different colors. 

% \subsubsection{Local Guide Interface}
% The local guide interface is similar to the user interface. The key difference lies in the dialogue components. There is a dialogue panel on the left part of the screen. It will display both the robot's response and dictations from the users in piled-up message bubbles, so the local experimenter can monitor the situation simultaneously.

% We implemented the user interface and robot interface using HTML/CSS and Vanilla JavaScript, and the backend server using Node.js. 

\subsection{System Implementation}
\paragraph{Robotic System}
We built the interface on the user end and the robot end as web applications hosted on Google Cloud. We utilized the Zoom Video SDK \footnote{Zoom Video SDK: \url{https://developers.zoom.us/docs/video-sdk/}} for the teleconferencing video call and used Double 3 SDK \footnote{Double 3 Robot SDK: \url{https://github.com/doublerobotics/d3-sdk.git}} for robot control and navigation. To extend the robot's ability to handle concurrent control from multiple users, we implemented the multi-user control mechanism using \textit{Locks}, allowing only one user to occupy the lock and send the control signal to the robot at a time. Once the user stops using the control, the lock was released and available to other users.

\paragraph{Agent Guide}

% The robot dialogue was powered by a Large Language Model (LLM) using GPT4 API \footnote{\url{https://openai.com/index/gpt-4/}} and controlled by an online experimenter through Wizard of Oz (WoZ) to correct misinformation from LLM hallucinations~\cite{achiam2023gpt}. The user's dictated speech was first transcribed into text using a speech-to-text module and fed into the GPT4 model. The generated response was further verified by the experimenter through an admin portal to correct any inaccurate information before sending it to the user. The robot's response was synthesized via a robotic voice using a text-to-speech module. The experimenter also triggered the robot to initiate a conversation with a list of prepared prompting questions. The topics of the prepared prompting questions included recalling memories of the place, comparing changes from the user's previous experience, and suggesting going to certain nearby attractions. 

\rr{The agent guide's response was either generated by a Large Language Model (LLM) using GPT4 API\footnote{GPT4 API: \url{https://openai.com/index/gpt-4/}} or created through Wizard of Oz (WoZ)  controlled by an online experimenter. To minimize response delays, we prepared a series of prompts that the experimenter could select and edit. The topics of the prepared prompting questions included recalling memories of the place, comparing changes from the user's previous experience, and suggesting going to certain nearby attractions. For LLM generation, the response was further verified by the experimenter through an admin portal to correct any inaccurate or hallucinated information~\cite{achiam2023gpt} before sending it to the user. On average, the agent guide's response time was 6.11 seconds, compared to the 2.51-second response time of the human guide. The full list of the prepared prompts is provided in the supplemental materials. }

 % \footnote{\url{https://developer.mozilla.org/en-US/docs/Web/API/SpeechRecognition}}
 % \footnote{\url{https://developer.mozilla.org/en-US/docs/Web/API/SpeechSynthesis}}
% Due to challenges in the robot's autonomous navigation in the wild, we used WoZ to control the robot when the user verbally instructed the robot to navigate to certain areas. 

% To receive guidance from the conversational agent, the users clicked the dictation button. Their speech was transcribed into text and sent to the server. The server used these messages to call the GPT4 API \footnote{https://openai.com/index/gpt-4/} to ask for answers. Then the answers were sent out to the admin, who decided whether a modification was needed. Apart from modifications, the admin is also able to send out pre-defined messages or write new messages to users. In this way, the malfunction and non-helpful responses by from the GPT4 API were greatly reduced.

\section{User Study}
We conducted a 2 $\times$ 2 (one \textit{versus} paired remote users; human \textit{versus} agent guide) between-subjects study. We denote the four study conditions as \textit{\textbf{C1}} (one user with the human guide), \textit{\textbf{C2}} (one user with the agent guide), \textit{\textbf{C3}} (two users with the human guide), and \textit{\textbf{C4}} (two users with the agent guide).

\subsection{Participants}
Participants were 38 alumni of the University of Wisconsin--Madison (17 women, 21 men, age 25--70 ($M=43.68$, $D=12.44$)) who graduated between 3--46 ($M=20.54, D=12.04$) years ago. They were recruited through the alumni association's member mailing list and online alumni groups from a professional social media platform.\footnote{LinkedIn: \url{https://www.linkedin.com}} Participants signed up for the study via an online screening survey that required participants to be alumni and 18 years or older. They were compensated with \$20 upon study completion.   

\subsection{Study Set up} 
Two experimenters, one local and one online, were present during the study. The local experimenter set up the robot in the field (\textit{i.e.}, the library or the lakefront alumni park). The online experimenter hosted a teleconferencing call with the remote participant(s) and the robot. In conditions where a remote user was accompanied by a human guide (C1 and C3), an undergraduate student or a university librarian served as the guide. The guide can provide information for the users and help the user drive the robot. The remote visit destination was assigned based on the participant's preference. Choices included two university libraries and the lakefront alumni park. In total, we had 10 sessions in the libraries and 19 sessions in the alumni park. Each study session lasted approximately one hour. All study protocols and materials have been approved by the institutional review board (IRB) of The University of Wisconsin--Madison.

 % shared in the screening questionnaire.
% The link to the teleconferencing call was shared with participants by email prior to the study. The video conferencing platform was developed by our research team as described in \S\ref{sec:system design}. 
% \todo{mention the case of the memorial library}
% an undergraduate student who is currently attending the university served as the guide during the alumni park visit. A librarian from the university library served as the guide during the library visit. 
% \todo{tasks for local guide?}

\subsection{Study Procedures}

\paragraph{Study Briefing}
Participants completed the consent form online prior to joining the teleconferencing meeting. After participants joined the call, the online experimenter first introduced the study agenda, the remote site, available routes, the robot's appearance and capabilities, the user interface, and the remote study setup. The participants were informed that the robot can avoid obstacles and that a local experimenter was monitoring the robot to handle any unexpected incidents. 
% \rr{The study briefing took approximately 10 minutes.}
% Then participants were asked to complete a pre-study survey.  
% Participants were informed that they can choose to show or hide their faces on the robot's screen by turning on or turning off their cameras. 

\paragraph{Robot Trial Session}
After the study briefing, the robot joined the same teleconferencing call. Participants then had a trial session to familiarize themselves with and practice using the robot. The experimenter instructed the participant to try each control button on the interface, including zoom in/out, tilt up/down, directional control buttons, and setting a destination point for the robot's semi-autonomous navigation. In agent guide conditions (C2 and C4), participants were instructed to try the dictation feature to engage in a dialogue with the robot. In the paired visitor conditions (C3 and C4), the participants took turns to try the interface controls. After the instruction, the participants were asked to interact with the robot on their own until they felt comfortable to start the remote visit session. \rr{The trial session took 10--15 minutes.} 

\paragraph{Remote Visit Session and Post-Study Interview}
Following the trial session, each participant interacted with the robot for approximately 20 minutes to explore their chosen location, talk to the guide or talk to local people. After the remote visit, the online experimenter rejoined the teleconferencing call and conducted a semi-structured interview with the participants. At the end, participants filled out a post-study survey to reflect on their experience and perception of the robot and provided demographic information. \rr{Between the training, trial and study sessions, participants interacted with the robot for a total of 25–-40 minutes. All participants visited the majority of the designated areas, \textit{i.e.}, areas shown in the map on the user interface, in the alumni park or in the libraries.}

% In the total duration of the one-hour study, the study briefing, the robot introduction, and the robot trial session took 15–-20 minutes; the experience took approximately 20 minutes; and the post-study questionnaire and interviews took 15--20 minutes. 
% In the conditions with a human guide, the robot when requested by the participants. The guide also introduced the participant(s) to bystanders who showed interests and explained the studies. In the conditions with the agent guide, participants engaged in dialogue with the robot, asking the robot or answering the robot's questions about the campus, and requesting the robot to navigate to certain locations. Participants were also prompted by the agent that they could stop and say hi if they wanted to talk to bystanders. 
% All participants encountered bystanders during the remote visit. Participants directly talked to bystanders when both participants and bystanders were willing to engage in conversation. 

% \paragraph{Interview and Post-study Survey}
% \todo{summarize the survey}
 % containing \xx seven-point Likert-scale questions for the robotde (C1 and C3), the guide joined the remote teleconferencing call with the user and the robot, engaged in dialogue with the remote participant(s) and control and the remote experience, 
% Trial session
% Remote visit session
% Interview and questionnaire

\subsection{Data Collection and Analysis}
All study sessions were video- and audio-recorded. We collected and analyzed three types of data: the interviews, the dialogue history during the remote visit, and the post-study survey. All data were provided in an osf repository \footnote{OSF: \url{https://osf.io/9y75q/?view_only=8a962f9dc9e74cc6a0dc831e7055707f}}. Below we describe the measures and analysis for each type of data. 

\paragraph{Interviews and Dialogue History} \label{sec-dialogue-analysis}
\rr{
We performed thematic analysis~\cite{braun2012thematic} for both the interview data and the dialogue history. The interviews and dialogue were first automatically transcribed using an online transcription tool \footnote{Automatic transcription tool: \url{www.otter.ai}} and then verified by a member of the research team. The first coder, who conducted all user study sessions and was familiar with the study data, independently coded all data. A second coder from the research team verified the codes. Disagreements in coding were discussed and resolved between the two coders and final codes were categorized into the main themes of the findings.} \rr{In addition, we performed quantitative analyses on the dialogue history to identify interlocutor types and dialogue durations for each topic. We calculated the duration of the following dialogue topics: ``\textit{Touring Information},'' ``\textit{Robot System Control},'' ``\textit{Past Experience and Place Connection},'' and ``\textit{Other Chat Topics}.'' The duration of each dialogue topic was presented as percentages of the duration of the interaction.
}

% The duration of each dialogue topic was calculated based on the start time and the end time of the speeches. The final duration was normalized by dividing the total interaction duration for each participant and was presented as percentages of the duration of the interaction. 

\begin{figure*}[!tb]
    \centering
    \includegraphics[width=\linewidth]{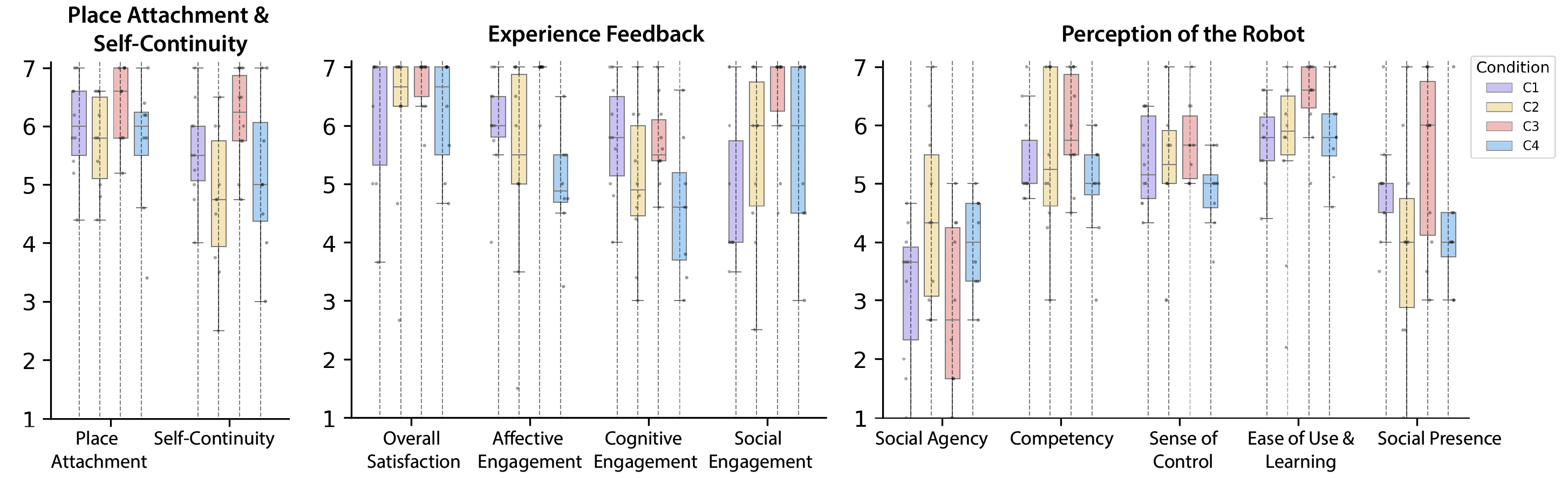}
    \caption{Results from our quantitative measures, including user's place attachment and sense of self-continuity after the remote experience (\textbf{\textit{Left}}), level of satisfaction, affective and cognitive engagement (\textit{\textbf{Middle}}), and perceptions of the robot (\textit{\textbf{Right}}). Participants reported an overall high level of place attachment and satisfaction of the experience. The sense of self-continuity, affective engagement and cognitive engagement were significantly higher when having the remote experience with the human guide comparing to an agent guide. The perceived robot social agency and user's social presence on the robot were significantly higher in the agent guide conditions. \textbf{\textit{C1}}: one user with the human guide; \textbf{\textit{C2}}: one user with the agent guide; \textbf{\textit{C3}}: two users with the human guide; \textbf{\textit{C4}}: two users with the agent guide.  
    }
    \label{fig:quant}
    \vspace{-9pt}
\end{figure*}

\paragraph{Survey}
% The full list of items are provided in the supplemental materials.
\renewcommand{\labelenumi}{\arabic{enumi}.}
\begin{table}[t]
\caption{\footnotesize{Place attachment and self-continuity measures. }}
\footnotesize
\centering
\label{tab: questionnaire items and loadings}
\begin{tabular}{p{0.9\linewidth}}
\toprule
\textbf{Place Attachment} \hfill Cronbach's $\alpha=0.83$ \\
% \midrule
\vspace{-4pt}
\begin{enumerate}[labelsep=1pt, itemsep=0.5pt, parsep=0pt, leftmargin=5pt]
   \item I feel at home during the remote experience.
    \item I am interested to learn about what is going on in UW--Madison.
    \item I want to know more about what is going on in UW--Madison after this experience.
    \item  If I had a chance, I would want to go back to UW--Madison to visit after this experience.
    \item This experience makes me feel attached to UW--Madison. 
\end{enumerate}
\vspace{-8pt}
\\

\midrule

\textbf{Perceived Self-continuity} \hfill Cronbach's $\alpha=0.86$ 
%  \\
% \midrule
% \vspace{-8pt}
 \begin{enumerate}[labelsep=1pt, itemsep=0.5pt, parsep=0pt, leftmargin=5pt]
    \item This experience connected me with my past.
    \item This experience connected with who I was in the past.
    \item This experience made me feel that there is continuity in my life.
    \item This experience made me feel that there is continuity between the past and the  present. 
\end{enumerate}
\vspace{-8pt}
\\
\bottomrule
\end{tabular}
\vspace{-16pt}
\end{table}

We administered a 55-item post-study survey. To understand the place-related experience and its psychological effect, we selected questionnaire items from a validated place attachment scale~\cite{raymond2010measurement}, self-continuity scale~\cite{sedikides2015nostalgia} and event experience scale~\cite{geus2016conceptualisation} and derived measures for place attachment, sense of self-continuity and level of affective and cognitive experiences. To understand user perception of the robot, we selected items from the USE questionnaire~\cite{lund2001measuring} and RoSAS~\cite{carpinella2017robotic} and constructed measures for the robot's social agency, competency, sense of control, ease of use and learning, and user's social presence. Table~\ref{tab: questionnaire items and loadings} lists questions for the measures of place attachment and perceived self-continuity. We report the questionnaire items for the measurements, items' reliability coefficients in the supplemental materials. A two-way ANOVA was performed to evaluate the effect of guide type (human \textit{versus} agent) and number of users (single \textit{versus} paired remote users) on all the measures.

% We performed factorial loading for the survey questions and constructed the following measures:
% sense of place attachment after the experience (five items, Cronbach's $\alpha=0.83$), sense of self-continuity after the experience (seven items, Cronbach's $\alpha=0.90$). 
% Feedback for the experience:
% overall experience (three items, Cronbach's $\alpha=0.78$), affective engagement (four items, Cronbach's $\alpha=0.86$), cognitive engagement (five items, Cronbach's $\alpha=0.85$), desire for bystander engagement (two items, Cronbach's $\alpha=0.88$). 

% We also constructed measures for the user's perception of the robot, including the robot's social agency (three items, Cronbach's $\alpha=0.80$), competency(four items, Cronbach's $\alpha=0.82$), sense of control(three items, Cronbach's $\alpha=0.71$), ease of use and learning (five items, Cronbach's $\alpha=0.84$), and user's social presence on the robot (two items, Cronbach's $\alpha=0.80$). 

 \begin{figure}[!b]
    \vspace{-12pt}
    \centering
    \includegraphics[width=\linewidth]{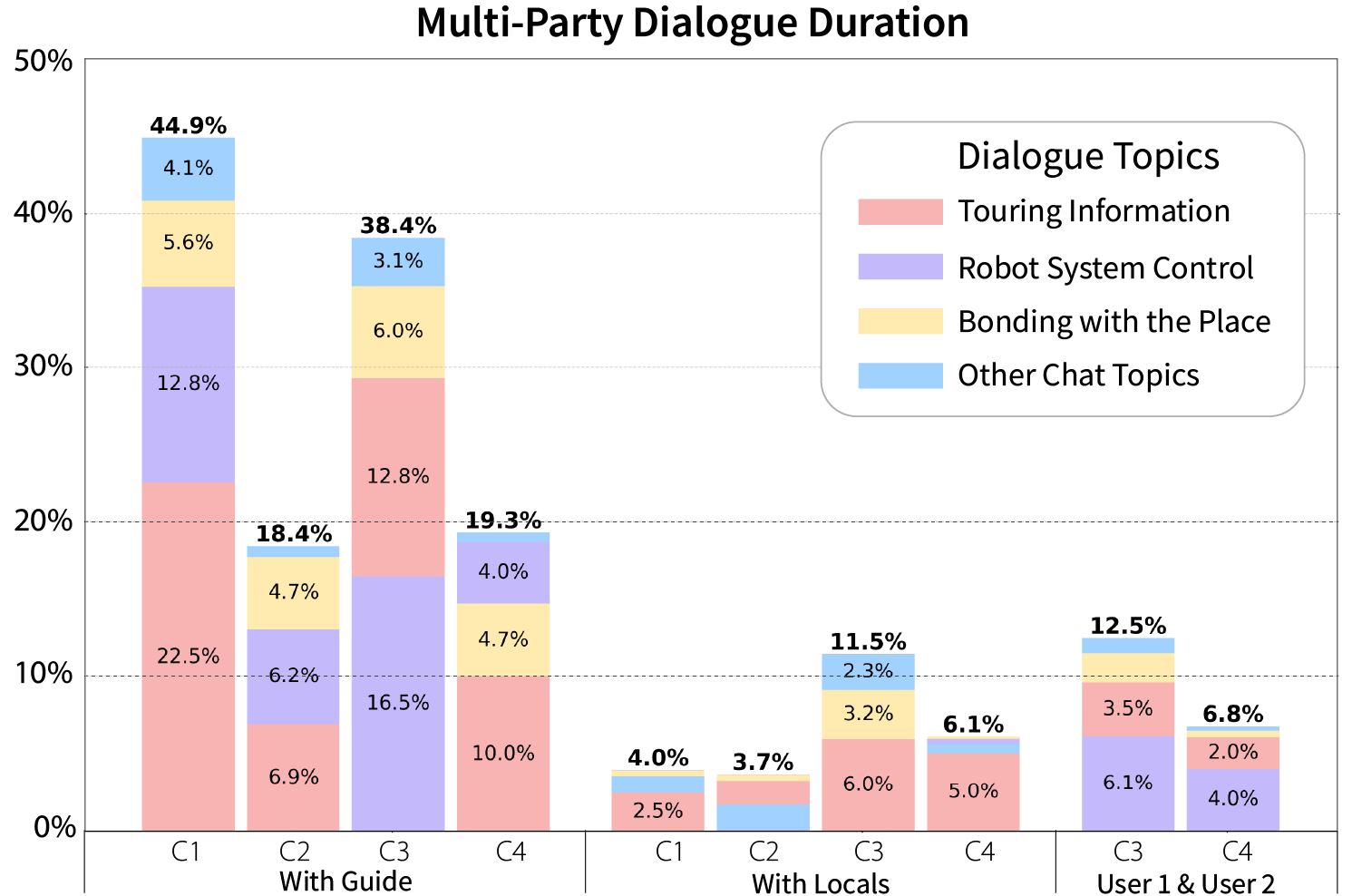}
    \caption{Data from our analysis of dialogue interactions including dialogues between user(s) and guide (\textit{\textbf{Left}}), user(s) and locals (\textit{\textbf{Middle}}), and within two remote users (\textit{\textbf{Right}}). Across all types of dialogue, participants engaged in significantly longer durations of dialogue when visiting with the human guide than visiting with the agent guide.}
    \label{fig:quant-dialogue duration}
    \vspace{-9pt}
\end{figure}

% \section{Results}
% Findings read like categorzation 

% Qualitative analysis --> more integrated, more synthesized ways of generating the findings
% --> initial analysis --> add a step --> describe the findings in the miro board --> how do we integrate into a story

% Integrate all findings 

% Present the findings combining everything 
% --> phenonemon --> may from an integration of all data

% \subsection{Quantitative Results}

% \subsubsection{Survey Results}

% \subsubsection{Interactive Behaviors Categorization}

% X-axis labels horizontal; text smaller
% Unsquish the text 
% Starting number is different 

% Group the categories - Bystanders -> Locals

\section{Findings}
% In the following section, we present our qualitative and quantitative findings on user experience with and use patterns of visiting places of meaning via a telepresence robot. 
Below, we refer to participants as ``users,'' the local guide as ``guide,'' and local bystanders and passersby as ``locals.'' 

% \todo{Statistical results: add captions, scales, figures clarity, fix label issues in the results}

\subsection{User Experience with and Perceptions of the Robot}

% We analyzed participants' overall satisfaction of the remote experience, level of affective engagement and cognitive engagement, and desire for bystander engagement across four conditions. We found a significant effect of visiting with a local guide on higher affective and cognitive engagement over visiting with the conversational agent. 
% A two-way ANOVA was performed to evaluate the effects of 

% The results indicated [no significant] main effect for [gender], F([1], [54]) = [0.50], p = [.483], partial η2 = [.01]; [no significant] main effect for [major], F([2], [54]) = [0.29], p = [.752], partial η2 = [.01]; and [a significant] interaction between [gender] and [major], F([2], [54]) = [3.71], p = [.031], partial η2 = [.12].

% \todo{Fix C3's visualization in Figure 3, and add another two or three sentences in the captions of Figure 3 \& 4 to summarize the statistical results (R1).}

% Figures \ref{fig:quant} and \ref{fig:quant-dialogue duration} show the data from survey responses and the user's dialogue engagement. Below, we report significant results from the analysis. The survey responses (Figure~\ref{fig:quant}) 

Survey results (see Figure~\ref{fig:quant}) indicate that the type of guide had a significant main effect on the user's sense of self-continuity ($F(1, 37) = 6.33, p = .017$), affective engagement ($F(1, 37) = 12.14, p = .0014$), and cognitive engagement ($F(1, 37) = 7.37, p = .010$); a marginal main effect on the user's place attachment ($F(1, 37) = 2.91, p = .097$). We also found a marginal main effect of the number of users on the desire for bystander engagement ($F(1, 37) = 4.28, p = .046$) and a marginal interaction effect between the guide type and number of users on the user's desire for social engagement ($F(3, 37) = 2.82, p = .10$). \rr{Simple effects tests show that the remote visit with local human guide significantly increased user's sense of self-continuity and affective and cognitive engagement, compared to visiting with the agent guide. The results also showed that visiting with another remote user increased the desire to interact with locals comparing to visiting alone.}

For user perceptions of the robot, we found a significant main effect of guide type on user perception of the robot's social agency ($F(1, 37) = 7.48, p = .0098$) and self-presentation of the robot ($F(1, 37) = 4.94, p = .033$). \rr{Participants perceived the robot as being more social, knowledgeable, compassionate, and felt more represented by the robot when visiting with the agent guide than visiting with the human guide.}

For dialogue engagement (Figure~\ref{fig:quant-dialogue duration}), the result revealed a significant main effect of guide types ($F(1, 37) = 15.22, p = .00064$) and a marginal main effect of number of users ($F(1, 37) = 3.51, p = .073$) \rr{on users' dialogue duration (including with the guide, local people, and other remote users). Participants were more engaged in dialogues when visiting with the local human guide than with the agent guide. Visiting with another remote user also increased users' overall dialogue engagement comparing to visiting alone.}
 
% \todo{Reviewers suggested improving our discussion of quotes from participants (R2, R3) and the depth of the qualitative findings (R3) in a way that is rooted in theory.
% We will extend our interpretations of the quotes by adding two or three sentences to explain the meanings of each subtheme and highlight their significance for understanding the place attachment experience (Sec V.B, C) and informing robot interaction design (Sec V.D). We will also add more detail to Figure 5 to improve the presentation of the qualitative findings (R2).
% }

\subsection{Place-Attachment Experience}
Participants reported experiencing nostalgia, happiness, excitement, and pride during the experience. They mentioned that seeing the campus brought back memories and evoked good feelings. \rr{In the following, we present users' place attachment experiences that are associated with individual, group, and community attachment experiences.}

\paragraph{\textbf{Individual Attachment Experience}}
Individual place attachment is associated with the memories of personal routines, iconic aspects of the environment, and changes overtime. Places of personal routines are often associated with mundane things, such as favorite study spots, places for spending leisure time, and familiar routes on campus. One participant experienced \inlineqq{good feelings}{C2P7} after seeing the stairs in front of a building because he used to go up these stairs to buy coffee every day. As he commented, \inlineqq{it was a satisfying feeling of being back there, because that was a ritual that I used to do frequently}{C2P7}. Another participant shared how seeing \inlineqq{small, little sections}{C3P5} brought back good memories: \inlineqq{Seeing the sun come down over Picnic Point reminded me of a paper I wrote in college for a geography class}{C3P5}. Individual attachment experience is also connected with the iconic things and views in the visit. Many participants mentioned the iconic ice cream and beers on campus when they were nearby the shops and restaurants. One participant asked the robot for ice cream \pid{(C4P4)}. Another participant \pid{(C4P8)} tried to buy beers from a bystander through the robot. Participants also responded positively to the iconic lake view and good weather, commenting that \textit{``Nothing beats a lakefront''}\pid{(C3P5)} and \textit{``It looks nice and sunny, same as I remember.''}\pid{(C2P10)}. Participants also highlighted that seeing how the campus changed brought back memories, \textit{e.g.}, commenting: \inlineqq{I got to remember my first experience on the union terrace as a student in summer 1977.}{C4P5}.

\paragraph{\textbf{Group Attachment Experience}}
\rr{
Group attachment is concerned with social ties in the place and is triggered by shared memories among friends and social groups. Going to the terrace by the lake reminded one participant about his friend, as he shared \textit{``I miss going to the terrace with my friends and playing basketball''} \pid{(C4P6)}. Another participant shared his time in the limnology center by the lake and asked the robot the cost of joining the sailing club: \textit{``How much does it cost to join Hoofers (outdoors club) these days?''} \pid{(C4P2)}. Participants \pid{(C1P5--6, C1P8, C3P3)} also recalled their socializing time, \textit{e.g.}, \pid{C3P3} mentioned a place for students to take breaks and socialize in between classes: \textit{``One of my vivid, vivid memories, was there was a kind of a casual space on the first floor where they continually played music... And that's where people [stutter] taking a break between one subject to the next.''}.} 

% P28 asked one bystander about the professor who taught him back in the days and 
% \paragraph{Experiencing Mundane Scenes}
% \pid{C1P5} shared that \inlineq{It's very important. Has one curate the experience. We have questions that can tell you what is, what you may want to know, the origin of something, or what's changed, or what's what's new, what's, what's available, or just tell me which way to go. I think that's that's clutch. I think it's great.}
% \paragraph{Social Encounter}
 % Below we report findings related to the community attachment including the desire to engage with the current students, see the statue of the school mascot and chat about the football games with local people.}
\paragraph{\textbf{Community Attachment Experience}}
Community attachment is associated with the sense of belonging to the community, the cultural identity of the place and the spirit of the place that elicits pride. Connection with current events on campus indicated community attachment. \pid{C3P5} shared his happiness when he got to talk to locals about the recent football game, saying, \inlineq{It again, connected the world... It was really neat. Everyone at least answered my question about Badger football and what they thought the season was going to be like.}. Participants also expressed their excitement seeing the statue of the school's mascot: \textit{``Hi Bucky (the mascot's name). Gotta stop and look at Bucky.''} \pid{(C1P1)}, \textit{``I've never seen this Bucky statue.''}\pid{(C3P2)}. Engagement with locals, particularly current students, also improved participants' sense of belonging to the community. Participants recalled similar things they used to do on campus:\inlineqq{To just see college students roaming around the library like, it felt like, you know, it felt like I was connected to campus}{C2P1}, \inlineqq{It helped me remember when I was young, and the experience there as well}{C2P7}. Three participants \pid{(C2P9; C3P9, 10)} even expressed envy when seeing students on campus and wished they were back. As \pid{C3P9} shared, \inlineq{Especially knowing that it was the first day of class, that's a that's a very fun day to be on campus, especially if you're a student. So I wouldn't call it jealousy, but something like that.}

\subsection{Use Patterns and Personas}
% \todo{some transitions about the thematic analysis and user interaction patterns and conclude the personas}

Four user personas emerged from our thematic analysis of the interview data and user interaction history, highlighting user preferences for interactions with the guide, locals and other remote users. 

% from our participants. The 

% We categorized users into the following four personas based on the interaction patterns of our participants.  

\subsubsection{Persona 1: \textbf{The ``Sightseer''}}
The first group of visitors saw their experience primarily as a ``sightseeing'' activity. They viewed social interaction with the guide or the locals as optional or unnecessary. They mentioned \inlineqq{driving the robot like that's more important}{C4P4} and the experience \inlineqq{was more just the sense of sight}{C3P10}. Sightseers did not actively engage with the agent guide. \pid{C2P2} ignored all questions from the agent, saying, \inlineq{I just didn't really feel like replying to the robot.} Comparing the remote robot experience with using Google Maps, she thought \inlineq{Google Maps would be maybe just as entertaining.} \pid{C2P7} felt strange that the robot asked about his memories, commenting, \inlineqq{Why would I share a memory with a robot}{C2P7}.
 % \pid{C1P9; C2P2, 7; C3P10; C4P4} 
% \pid{C1P9} only provided very brief responses to the guide's questions and commented, \inlineq{I was more focused on, like, controlling the robot than necessarily, like chatting with the ambassador.} 

% \inlineqq{Why do you care? You're a robot}{C2P9} and 

% \begin{figure}[!tb]
%     \centering
%     \includegraphics[width=\linewidth]{Figures/persona.png}
%     \caption{Four personas identified in our analysis of use patterns.}
%     \label{fig:persona}
% \end{figure}

% Participants in this category were mainly under conversational agent conditions (\xx). 
% Their dialogue with the robot was focused on controlling the robot or seeking factual information about the university. 
\subsubsection{Persona 2: \textbf{The ``Tourist''}}
This type of visitor was the information seeker, who engaged in dialogue with the guide. Participants both gave the robot direction (\textit{e.g.}, \inlineqq{Robot let's go back to the window}{C2P5}) and requested information about the history of the university, the number of visitors each year, name of the building in view, price of a football game, etc. During dialogue with the guide, tourists' main goal was to get information about the place or to receive support to navigate the robot. They did not actively engage in social interaction with the locals and often chose to turn off their cameras and hide their face on the robot. For example, \pid{C1P10} felt uncomfortable talking with strangers, saying that \inlineq{I still feel uncomfortable about talking to like random people in the public, just because I feel like I'm bothering them, but I thought having a guide there to interact with was helpful}.

\subsubsection{Persona 3: \textbf{The ``Socialite''}}

The third type of user saw social interaction as the main goal of the experience, engaged socially with the guide, and actively sought interaction with locals. They shared with locals details of their personal connection with the university, explained the current tour, and asked them about ongoing campus events. Participants commented, \inlineqq{The nature of this is kind of a social interaction}{C3P3} and \inlineqq{the interaction with other people and their reaction to the robot was the highlight}{C2P6}. To them, engagement with locals created a \inlineqq{true experience}{C1P9}, which made them feel \inlineqq{more alive}{C1P10}. Interacting with locals also improved their sense of presence, leaving them feeling \inlineqq{actually there}{C3P7} and \inlineqq{right there in the building}{C1P8}. \pid{C4P6} coincidentally met a friend who was working on campus. The encounter with the friend elevated the excitement in the experience: \inlineqq{That made it really fun. Yes, I think seeing somebody you know, like, enhanced it so much}{C4P6}. The robot created unique opportunities to interact with strangers. Participants \pid{(C3P1, 2, 6)} reported that they would not have interacted with locals if it had not been through the robot. As \pid{C3P6} shared, \inlineq{I would not have gone up to these people and said hello to them, had it not been a robot experience.}

\subsubsection{Persona 4: \textbf{The ``Companion''}}
The last type of visitors were participants who valued sharing the remote experience with the other remote users. Among our paired participants, one pair was a married alumni couple and another was an alumni father and son. For the father-son pair, the son \pid{C4P8} found the experience fun because of his father \pid{C4P7} was \inlineq{joking around with people.} The married couple shared that the visit experience evoked shared memories and expected to continue to discuss it after the study session, commenting that \inlineqq{After we're done here, he's going to come to me and say, hey, you know that was really neat. You remember where we did this and where we did that? So it was neat doing it together}{C3P6}. Although participating on their own, four participants \pid{(C2P1, 8; C3P1, 2)} wished to have this experience with family members or friends who are also alumni in the future. One participant \pid{(C2P8)} took a picture of the robot's view of the campus and shared it in a group chat with his alumni friends. Another participant wanted to have this experience again with his wife whom he proposed at the lakefront park: \inlineqq{I asked her to marry me on the terrace, I would have loved to do that whole experience with her and to look at things together}{C3P2}.

% Another participant \pid{(C3P1)} mentioned that this type of remote experience could be a ``family affair''\pid{(C3P1)} in the future. 
% He commented, \inlineqq{It's making me laugh. So, and you can talk about, you know, what you're doing with somebody else. Instead of just, I think if it was you by yourself, you know, rolling around, it could get boring}{C4P8}.

% \paragraph{Future Use for Family and Friends Sharing the Experience} 
% [Names of Campus Attractions] = the terrace and the lake

% The participants also expressed the desire to have this experience together with friends who live nearby, saying that \inlineqq{I could see sitting with a couple of them and using the robot to look at sites and talk to people. That would be an absolute blast}{C3P2} and \inlineqq{Maybe we're, like, sitting watching a football game, and at halftime, we like, could do a little tour and reminisce about our time on campus}{C2P1}. 
% [Name of Campus Attraction] = on the terrace

% \textbf{Share the experience with other friends}
% C2P8: I was actually did a WhatsApp, we have a group, and I did a screenshot in a *** of WhatsApp, or Whatsapp group, like, what am I viewing now?

\subsection{Multi-Party Interactions}

% \subsection{Interactions with Locals and Other Users}

% \subsubsection{Spontaneous Social Engagement}
In the following section, we present the interaction patterns and challenges of the interactions with locals, other remote users, and the guide.

% Some participants approached locals and tried to initiate conversation, and sometimes locals approached the robot and started talking to remote users. 

% Similarly, interrupting other people was not a concern for \pid{C3P5} due to the novelty of the technology: \inlineq{I thought it was actually easier because you can kind of like, kind of like interrupt people, and you're not quite interrupting them, because it's such a unique thing that they want to see what's going on.}

 % Similarly, \pid{C3P1} and \pid{C3P2} commented that it would be \inlineq{creepy} if it was in person. 
% \subsubsection{Challenges in Engagement with Locals}
% Unfriendly bystanders 
% Psychological safety
% Robot safety 

% Several participants (\xx) were ignored by locals when they tried to initiate a conversation, 
% as illustrated in the exchange below: 

% \begin{quote}
% \pid{C2P10}: \inlineq{Hey, Guys.} [after 8s] \inlineq{Hello.} [after 15s] \inlineq{Hello, hello.} [after one minute] \inlineq{Can you talk to me for a second? Hello. Hello.} [after two minutes] \inlineq{Hello, can you talk for a minute?}
% \end{quote}
\paragraph{Interacting with Locals}
The main trigger for interaction with locals appeared to be the novelty of the robot. Locals approached the robot out of curiosity, or participants wanted to introduce the robot to locals \pid{(C1P3--4, C3P1--6)}. As \pid{C1P4} shared \inlineq{[They] started talking to me because they were curious about what was going on, what this was about.} On the other hand, participants \pid{(C2P8, 9; C3P2)} reported challenges in initiating conversations, not knowing who might be interested in talking to the robot  \pid{(C2P9)} and feeling \inlineq{awkward} with \inlineqq{some of the looks the robot got}{C1P10}. 11 participants, mostly in the agent conditions, tried to greet locals, but no one responded \pid{(C1P9; C2P1, 4, 6, 7, 10; C3P5, 7;  C4P3, 6, 8)}. Some locals became friendly only after knowing that remote users were alumni. As \pid{C3P9} commented, \inlineq{I thought it was interesting also to see how people responded, because people seem to be a little bit unwelcoming until we were introduced.} A few participants encountered locals who were unfriendly. A local who was a security guard questioned the legitimate operation of the robot, asking \inlineq{Should this be here?} These unfriendly reactions surprised \pid{C2P8} who commented \inlineq{The coldness of the people, or like, the sort of strange comments or disinterest or whatever you know, that kind of turned me off a little bit.} Given the challenges of engagement with locals, the participants suggested various ways to increase locals' awareness of the robot and the remote users. \pid{C3P4} suggested attaching a sign to the robot \inlineq{as awareness of a tour in progress} for locals. \pid{C3P5} further suggested the use of an attachment, such as a red bell with a \inlineq{Come to meet some alumni} sign, to \inlineq{encourage people to engage.}

\subsubsection{Interacting with Remote Users}

% \subsubsection{Control Delegation and Collaboration}
Paired participants took turns controlling the robot. Participants shared that the ability to relegate control of the robot to their partner gave them time to do other things. \pid{C3P5--6}, who received guidance from the conversational agent, described how the shared control left them time to engage in the dialogue with the robot: \inlineqq{When I wasn't controlling it, I could think of more questions because I wasn't distracted by controlling the movement.}{C3P5} Another participant mentioned that sharing the experience helped mitigate control challenges: \inlineqq{I liked it better having somebody to kind of do it with... could talk with them and not feel silly when you made mistakes on navigating}{C3P7}. Furthermore, participants (\pid{C3P1, 8; C4P1}) shared that they could learn from each other in using this new technology and help each other. \pid{C4P1} liked the \inlineq{teamwork aspect} and shared that \inlineq{I both are kind of like getting used to this new technology and kind of like figuring out together.} Similarly, \pid{C3P8} highlighted the value of learning from observation, stating, \inlineq{taking turns driving, we kind of saw each what each other did, and observed and learned from that.}

\subsubsection{Interacting with Guides}
The main dialogue topics with the guides included the robot control, information about the campus, and personal bonding with the university as illustrated in Figure~\ref{fig:quant-dialogue duration}. 
Participants appreciated that the local human guide could \inlineqq{curate the experience}{C1P5}, which would \inlineqq{provide context}{C3P4}, and tell them \inlineqq{what typically happens there, what students do, or what kind of events are going on}{C1P10}. The local guide also helped participants connect with locals, \textit{e.g.}, introduced the robot and invited locals to talk to the remote users. \pid{C3P9} emphasized the need for a guide: \inlineq{You almost need the other [guide] explain what's going on before you start that conversation.} In agent guide conditions, several participants tried to be playful with the robot and teasing the robot. \pid{C2P10} asked the robot about a TV show, \inlineq{Did you ever watch Big Bang Theory? In one episode, Sheldon made a robot that appears to be similar to you.} Three participants asked the robot \inlineqq{Do you enjoy being a robot}{C2P9}, \inlineqq{What's your favorite kind of ice cream}{C4P6}, and \inlineqq{Did they ever give you robot cheese curds}{C4P7}. Participants highlighted how the dialogue improved engagement and made the experience more \inlineqq{interactive}{C4P2}.

\section{Discussion}
% \todo{Connect discussion with literature: what are the new points to be added? what are the literature to be added? findings - extension - literature}

% \todo{
% add several sentences to the first paragraph of Sec VI to root the discussion on user preferences and patterns in theory by highlighting the importance of social engagement, echoing that social bonding and community bonding are essential for place attachment (Low et al., 1992, Hidalgo et al., 2001).
% }

% Key discussion points:

% - Social engagement is important for the connection to the present
% -- Facilitate social interactions - motivations of the bystanders
% -- Playfulness of the experience 
% -- Social presence of the user / embodiment of the robot
% -- Eembodiment of the experience 
% -- Bystanders' perception of the robot
% Place attachment is concerned with changes over time, and the dialectic of now and then~\cite{lomas2024imagined}. 

\rr{
Telepresence robots provide unique opportunities for users to experience a remote place of personal significance. Our participants reported place-related experiences at the personal, group, and community levels. The robot connected people with the place from their past, triggering personal memories about mundane daily routines, social experiences, and life milestones. The robot also linked people with the place in its present, engaged them in the live community, helped them experience the landmarks of the place, and elicited a sense of pride and bonding with the place.} In answering our $RQ_1$, \textit{How do individuals or groups use telepresence robots to connect with a place,} we highlighted place attachment experiences on the personal, group, and community level and four personas that exhibit different interaction styles with the robot for place engagement. In answering  our $RQ_2$, \textit{How can telepresence robots facilitate human-place connection,} we reported users' interaction patterns and challenges during the experience, including use patterns when robot control was shared with other online users and challenges in social engagement with local people. Below, we discuss the design implications of our findings, focusing on ways to support interactions with local users and to support multiple robot users.

\subsection{Design Implications}
% Our findings highlighted four different personas for the experience, pointing to the needs for customization on the guide types and companions. 

\subsubsection{Designing for Interactions with Locals}
\rr{
Social ties and community engagement are essential to forming and maintaining place attachment~\cite{hidalgo2001place, low1992place}. Our findings indicated that interacting with people through the telepresence robot, including the human guide, other online users, and local people in the environment, enhanced users' sense of presence and engagement and reconnected them with the community. The remote robot provided rich opportunities for users to act in the place, \textit{e.g.}, engaging in events and interacting with locals. Nevertheless, we observed several challenges related to social engagement, \textit{e.g.}, encountering bystanders unwilling to engage and difficulty in initiating conversations with locals. These findings highlight the importance of locals' awareness of the robot and the remote user's environmental awareness~\cite{heshmat2018geocaching, li2024teleaware}. Future design of the robot can facilitate social engagement between the remote user and locals. For example, one of our participants suggested attaching a sign on the robot that signaled who is connected with the robot, such as \textit{``Come to chat with alumni.''} The robot may detect unwillingness to interact and intervene politely~\cite{chin2020empathy, hu2022polite}, \textit{e.g.}, by saying ``\textit{Sorry for the interruption, let's proceed to the next destination.}''}

\subsubsection{Designing for Interactions with Multiple Robot Users}
\rr{
Distinct collaborative control behaviors emerged when multiple users accessed the robot, including taking turns to control the robot, delegating the control to one user, and helping and learning from each other. Prior work on multi-user robotic systems has emphasized the need for information awareness in designing user interfaces~\cite{rutha2021role}. The need to support place experience creates new challenges, as users can have different preferences for things to see and people to talk to in the remote place. Critical considerations include creating awareness~\cite{drury2003awareness} for other remote users and ensuring information transparency in designing the multi-user robot interface. Our findings also highlighted participants' stronger desire to engage with bystanders visiting with another remote user compared to visiting alone and revealed playful interactions among them. Future research can investigate the social dynamics among multiple robot users and the effects of group factors on playfulness in the remote experience.}  
% , the robot's autonomy~\cite{beer2014toward} and the remote environment, 
\subsubsection{Designing to Promote Place-Related Experiences}
% under different weather conditions, navigates across various terrains, 
% \paragraph{Environmental Factors}
We focused on university alumni's place attachment to their \textit{alma mater}, while future work can be expanded to other types of places and elicit different forms of attachment. For example, the destinations can also be symbolic places that carry significant meaning for them~\cite{russell2012remembering}, such as sacred or historically significant places, or places associated with ``imagined communities''~\cite{phillips2002imagined} such as a country whose language an individual is learning as a second language~\cite{liao2018exploring}. Our study locations included an outdoor environment where multiple environmental factors such as weather, physical characteristics of the location (\textit{e.g.}, long distances), and crowds affected the remote experience. Our participants expressed interest in visiting various campus locations that they felt attached to, but it was time-consuming or impossible to drive the robot to these locations due to long distances or lack of accessibility. The future designs can explore how multiple robots placed at key locations can allow remote visitors to switch views without driving the robot between locations. It is also critical to engage community stakeholders such as facility administrators and policymakers to designate areas to deploy these robots in ways that are safe and do not disrupt the environment~\cite{kayukawa2022users}. In our study, we observed the need to provide background or historical information about the place, but guides did not always have this information. Future designs may consider utilizing historical records about locations or obtaining oral history by asking for help from locals. 

   % 
  
% \paragraph{Information Support}
% Existing literature have studied the use of robots to provide guidance in various touring contexts~\cite{hu2024really, ng_cloud_2015, burgard1999experiences, thrun1999minerva}, 
% \paragraph{Future Places}

% \todo{Finally, we appreciate R2's suggestions to deepen the discussion, including the bystander perceptions and interactions with the robot, embodiment and social presence of the robot, motivators of the visitors to return to the place, multi-user relationships and dynamics, exploration of playfulness during the experience. We will add these points to Section VI as space allows.
% }

\subsection{Limitations \& Future Work}
% \todo{R3 pointed out the novelty effect of using the robot as a potential limitation. Between the training and the trial and study sessions, participants interacted with the robot for a total of 25–40 minutes. We acknowledge that this short period may not be sufficient to overcome novelty effects, which may affect user experience. We will include this point in our discussion of study limitations.
% Although the latency of the virtual guide is longer than the human guide, our observations suggest that this time is within a tolerable range. We will specify the latency time in the System Design Section and acknowledge it in the Limitations Section.
% }

Our robotic platform was unable to provide consistent high-quality video due to connectivity and bandwidth limitations. Responses from the agent guide involved a noticeable delay, which might have negatively affected user experience. Finally, future work should investigate long-term human robot interactions to understand the potential impact of novelty effects. 

\section{Conclusion}
We explored the design space of using telepresence robots to support place attachment. Our study investigated multi-party telepresence robot interactions and characterized robot-facilitated place-attachment experiences on the personal, group, and community levels. We proposed design implications for place-related experiences and suggested robot design facilitating social participation and multi-user interactions. 
% - use more autonomous system rather than wizard of oz; utilize the LLM more
% - novelty effect of bystanders, what if people are not interested
% - limitation of the robot system, quality of the video and navigational stability - affected by the sun
% - more pair who know each other 
% - better control for the user's self-presence (mute or unmute)

\section*{Acknowledgments}
This work received support from the \textit{McPherson Eye Research Institute ``Expanding Our Vision Award''} and \textit{Google Award for Inclusion Research Program}. Figure 1 modified images by Freepik for its design. We thank the \textit{Wisconsin Alumni Association} for help with recruitment, Irene Ho for help with the teaser figure and our participants for their time and feedback. 
% The preferred spelling of the word ``acknowledgment'' in America is without 
% an ``e'' after the ``g''. Avoid the stilted expression ``one of us (R. B. 
% G.) thanks $\ldots$''. Instead, try ``R. B. G. thanks$\ldots$''. Put sponsor 
% acknowledgments in the unnumbered footnote on the first page.
\balance
\bibliographystyle{IEEEtranN}
\bibliography{IEEEabrv,references.bib}

\end{document}